\documentstyle[preprint,aps,eqsecnum]{revtex}

\def\beq#1{\begin{equation} \label{#1}}
\def\eeq{\end{equation}}
\newcommand{\bea}{\begin{eqnarray}}
\newcommand{\eea}{\end{eqnarray}}
\def\bra#1{\left\langle #1\right\vert}
\def\ket#1{\left\vert #1\right\rangle}
\def\epsp{\epsilon^{\prime}}
\def\NPB{{ Nucl. Phys.} B}
\def\PLB{{ Phys. Lett.} B}
\def\PRL{ Phys. Rev. Lett.}
\def\PRD{{ Phys. Rev.} D}

\begin{document}
{
\tighten
%\preprint {\vbox{
% \hbox{WIS/07/00-Apr.DPP}
% \hbox{TAUP 2629-2000}
% \hbox{hep-ph/}
% \hbox{ANL-}
%}}

\title {Quantum Mechanics of Neutrino Detectors Determine
Coherence and Phases in Oscillation Experiments}
\author{Harry J. Lipkin\,\thanks{Supported
in part by grant from US-Israel Bi-National Science Foundation
and by the U.S. Department
of Energy, Division of High Energy Physics, Contract W-31-109-ENG-38.}}
\address{ \vbox{\vskip 0.truecm}
  Department of Particle Physics
  Weizmann Institute of Science, Rehovot 76100, Israel \\
\vbox{\vskip 0.truecm}
School of Physics and Astronomy,
Raymond and Beverly Sackler Faculty of Exact Sciences,
Tel Aviv University, Tel Aviv, Israel  \\
\vbox{\vskip 0.truecm}
High Energy Physics Division, Argonne National Laboratory,
Argonne, IL 60439-4815, USA\\
~\\harry.lipkin@weizmann.ac.il
\\~\\
}

\maketitle

\begin{abstract}
The apparent symmetry between energy and momentum found in
all covariant descriptions of neutrino oscillations is destroyed in the
neutrino detector, a quantum mechanical system described by a density matrix
diagonal in energy but not in momentum. The off diagonal matrix elements
between states of different momenta and the same energy  produce the coherence
and interference between mass eigenstates having the same energy and different
momenta that produce oscillations.
\end{abstract}

} % end tighten

% my definitions

\def\beq#1{\begin{equation} \label{#1}}
\def\eeq{\end{equation}}
\def\bra#1{\left\langle #1\right\vert}
\def\ket#1{\left\vert #1\right\rangle}
\def\epsp{\epsilon^{\prime}}
\def\NPB{{ Nucl. Phys.} B}
\def\PLB{{ Phys. Lett.} B}
\def\PRL{ Phys. Rev. Lett.}
\def\PRD{{ Phys. Rev.} D}

The continuing argument about the roles of energy and momentum in neutrino
oscillations has been resolved by the observation that all neutrino experiments
involve detectors which are quantum mechanical systems at rest in the
laboratory system and whose quantum mechanics play a crucial role\cite{neusb}.
This point is clearly overlooked in a recent paper\cite{Giunti} which
criticizes the ``equal energy assumption" of another paper\cite{Okun} by
focusing only on the properties of the neutrino wave packet traveling between
source and detector and completely ignoring the quantum mechanics of the
detector.  The so-called ``equal energy assumption\cite{Okun}" has been shown
previously to arise  naturally from the interaction of the neutrino with its
environment\cite{Leo}, but has been questioned because of its ``stationarity"
assumption. The loophole in Stodolsky's argument that  experiments measuring
time can violate stationarity has now been closed by  a rigorous
quantum-mechanical calculation\cite{neusb} of the detection process that does
not assume stationarity.

In any realistic experiment the neutrino  wave packet is detected  by a
quantum-mechanical detector which recognizes coherence between neutrino
amplitudes with the same energy and different momenta. The coherence and
relative phases between components of the wave packet with different energies
are either destroyed in the detector or rendered irrelevant to the flavor
spectrum of the outgoing charged leptons.

The detection of the neutrino is a weak interaction described in first order
perturbation theory by transition matrix elements between the initial state of
the neutrino-detector system before the interaction and all possible final
states. The initial state of the detector is described in the laboratory system
by a  density matrix which is diagonal in energy but not in momentum. It is
shown in ref.\cite{neusb} that the off-diagonal elements in momentum of the
density matrix of the initial state of the detector determine coherence and
phases of neutrino oscillations.

This asymmetry between energy and momentum in the initial detector state
destroys the apparent symmetry between energy and momentum noted in all
covariant descriptions of neutrino oscillations. A fully covariant description
of any experiment which can be used also to consider detectors moving with
relativistic velocities is not feasible at present. A covariant description
which neglects the quantum mechanics of the neutrino-detector interaction is
neglecting some essential physics of all realistic oscillation experiments.

The simple hand-waving argument for this physics states that the uncertainty
principle and the localization in space of the detector nucleon that absorbs
the neutrino prevents the detector from knowing the difference between
components of the incident neutrino wave packet with slightly different momenta
and the same energy.

The rigorous quantum-mechanical argument notes that the product  $\delta p
\cdot \delta x$ of the quantum fluctuations in the position of the detector
nucleon $\delta x$ and the range of momenta $\delta p$ in relevant neutrino
states having the same energy is a small quantity. Taking the leading terms in
the expansion of the transition matrix elements in powers of $\delta p \cdot
\delta x$ gives the result that the flavor spectrum of the charged leptons
emitted from the detector at a given energy is determined by the relative phase
of the components of the incident neutrino wave packet having the same energy
and different momenta.

At this stage time measurements and all possible coherence between amplitudes
from  components of the incident neutrino wave packet with different energies
are not considered and unnecessary. There may be fancy time measurements which
can introduce such coherence. But the coherence between incident neutrino
states with the same energy and different momentum already determines the
flavor output of the detector for each incident neutrino energy and cannot be
destroyed by  time measurements.

There remains the question of the possible variation of flavor output of the
detector as a function of energy. As long as this flavor output does not change
appreciably over the relevant energy range in the wave packet, the standard
neutrino oscillation formulas are valid. When the flavor output varies widely
as a function of energy, oscillations are no longer observed. This can be seen
in the case of neutrinos traveling large distances with many oscillation wave
lengths, as in neutrinos arriving from a supernova. Here the neutrino wave
packet separates into components with different mass eigenstates, traveling
with different velocities and reaching the detector at measurably different
times. All this time variation appears simply\cite{Leo} in the energy spectrum,
which is the  fourier transform of the time behavior.

\section{acknowledgments}

It is a pleasure to thank Boris Kayser, Lev Okun, Leo Stodolsky and Lincoln
Wolfenstein for helpful discussions and comments.

\end{document}